\documentclass[superscriptaddress, twocolumn,amsmath,amssymb,prb]{revtex4-1}
\usepackage{graphicx}
\usepackage{dcolumn}
\usepackage{multirow}
\usepackage{booktabs}
\usepackage{bm,color}

\begin{document}

\title{Strong, anisotropic anomalous Hall effect and spin Hall effect in chiral antiferromagnetic compounds
Mn$_3X$ ($X$ = Ge, Sn, Ga, Ir, Rh and Pt)}

\author{Yang Zhang}
\affiliation{Max Planck Institute for Chemical Physics of Solids, 01187 Dresden, Germany}
\affiliation{Leibniz Institute for Solid State and Materials Research, 01069 Dresden, Germany}

\author{Yan Sun}
\affiliation{Max Planck Institute for Chemical Physics of Solids, 01187 Dresden, Germany}

\author{Hao Yang}
\affiliation{Max Planck Institute for Chemical Physics of Solids, 01187 Dresden, Germany}
\affiliation{Max Planck Institute of Microstructure Physics, Weinberg 2, 06120 Halle, Germany}

\author{Jakub \v{Z}elezn{\'y}}
\affiliation{Max Planck Institute for Chemical Physics of Solids, 01187 Dresden, Germany}

\author{Stuart P. P. Parkin}
\affiliation{Max Planck Institute of Microstructure Physics, Weinberg 2, 06120 Halle, Germany}

\author{Claudia Felser}
\affiliation{Max Planck Institute for Chemical Physics of Solids, 01187 Dresden, Germany}

\author{Binghai Yan}
\email{yan@cpfs.mpg.de}
\affiliation{Max Planck Institute for Chemical Physics of Solids, 01187 Dresden, Germany}
\affiliation{Max Planck Institute for the Physics of Complex Systems, 01187 Dresden,Germany}

\begin{abstract}

We have carried out a comprehensive study of the intrinsic anomalous Hall effect
and spin Hall effect of several chiral antiferromagnetic compounds,
Mn$_3X$ ($X$ = Ge, Sn, Ga, Ir, Rh and Pt) by $ab~initio$ band structure and Berry phase calculations.  These studies reveal large and anisotropic values of both the intrinsic anomalous Hall effect and spin Hall effect.
The Mn$_3X$ materials exhibit a non-collinear antiferromagnetic order which,
to avoid geometrical frustration, forms planes of Mn moments that are arranged
in a Kagome-type lattice. With respect to these Kagome planes, we find that both the anomalous Hall conductivity (AHC) and the spin Hall conductivity (SHC) are quite anisotropic for any of these materials.
Based on our calculations, we propose how to maximize AHC and SHC for different materials.
The band structures and corresponding electron filling, that we show are essential to determine the AHC and SHC, are compared for these different compounds.
We point out that Mn$_3$Ga shows a large SHC of about 600
$(\hbar/e)(\Omega\cdot cm)^{-1}$.
Our work provides insights into the realization of strong anomalous Hall effects and spin Hall effects in chiral antiferromagetic materials.

\end{abstract}


\maketitle

\section{introduction}\label{intro}

\begin{figure}[htbp]
\begin{center}
\includegraphics[width=0.5\textwidth]{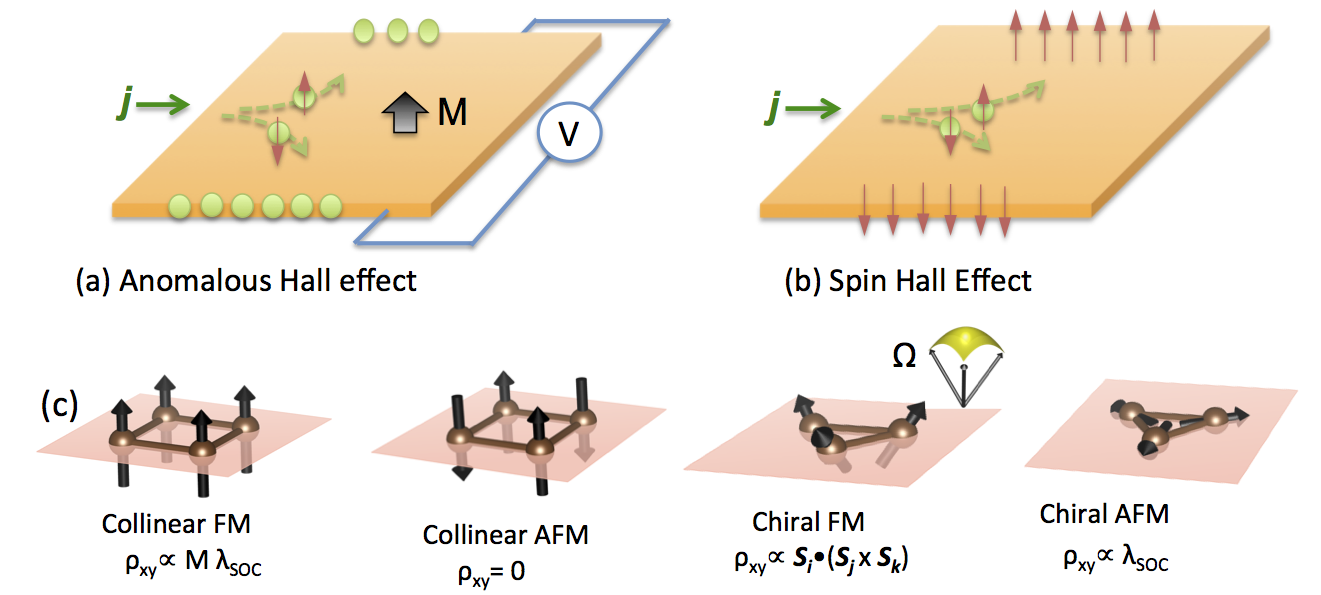}
\end{center}
\caption{
Schematic illustrations of (a) the anomalous Hall effect and (b) the spin Hall effect from the viewpoint of spin-dependent Mott scattering. (c) The anomalous Hall effect in collinear FM, collinear AFM, chiral FM, and chiral AFM systems. $\rho_{xy}$, $M$, $\lambda_{SOC}$ and $S_i \cdot (S_j \times S_k)$ represent the anomalous Hall resistivity, magnetization, strength of SOC and the scalar spin chirality, respectively.
}
\label{AHESHE}
\end{figure}

The anomalous Hall effect (AHE)~\cite{Nagaosa2010} and spin Hall effect
(SHE)~\cite{Sinova2015} are very important members of the family of Hall
effects. The AHE is characterized by a transverse voltage generated by a
longitudinal charge current usually in a ferromagnetic (FM) metal.  The AHE can
be generalized to the case of the SHE in nonmagnetic materials in which Mott
scattering~\cite{Dyakonov1971} leads to the deflection of spin-up and -down
charge carriers in opposite directions, owing to spin-orbit coupling (SOC), as
illustrated in Fig. ~\ref{AHESHE}. Thus, a longitudinal charge current can
generate opposite spin accumulations along opposing edges in the transverse
direction to the current. On the contrary, a spin current can also induce a
transverse voltage drop, in an effect called the inverse SHE.
Both the AHE and SHE are of particular interest for spintronic
applications~\cite[and references therein]{Jungwirth2012,Maekawa2012,Hoffmann2013}
in which spin currents can be used to manipulate magnetic moments, for example,
switching the state of magnetization of magnetic nano-elements, or for inducing
the very efficient motion of domain walls~\cite{Parkin2015,Yang2015}.
Thus, the SHE has recently attracted much attention by both experimentalists
and theorists, and there has been widespread efforts to search for candidate materials that exhibit strong AHE or SHE.

The AHE and SHE originate from the electronic and magnetic structures of materials and have both extrinsic and intrinsic origins. Extrinsic contributions depend sensitively on impurity scattering while intrinsic effects are derived from properties of the band structure.  It is the intrinsic AHE and SHE that are the subject of this article.
For the AHE of an ordinary collinear ferromagnet, it has been established that the Berry curvature, a quantity closely determined by the band structure, acts as a fictitious magnetic field in momentum space, that is derived from the magnetization and SOC, and affects the charge motion in the same way as a real magnetic field~\cite{Xiao2010}.
In a collinear AFM, it is not surprising that the AHE vanishes due to the spin-up and -down conduction electron symmetry, or rather the existence of a symmetry by combining a time-reversal symmetry operation and a lattice translation.
In a chiral ferromagnet where magnetic moments are tilted in a lattice, it was
recently found that the aforementioned fictitious magnetic field can also be generated by the scalar spin chirality~\cite{Ohgushi2000,Taguchi2001}, $S_i \cdot (S_j \times S_k)$ ($S_{i,j,k}$ denote three non-coplanar spins), which does not necessarily involve SOC. When an electron makes a loop trajectory in a chiral FM lattice, the electron acquires a real-space Berry phase due to double exchange interactions with the chiral lattice spins. The corresponding AHE has been referred to as a so-called real-space topological Hall effect in the literature (e.g.~\cite{Neubauer2009}).
In a chiral AFM in which the magnetic moments are coplanar, the topological Hall effect disappears because of the zero spin chirality.
However, an AHE can still exist due to a nonzero Berry curvature induced by the SOC~\cite{Chen2014Mn3Ir}. Indeed, a strong AHE was recently observed in the chiral AFM compounds Mn$_3$Sn and Mn$_3$Ge~\cite{Chen2014Mn3Ir,Kubler2014,Nakatsuji2016,Nayak2016}.
In principle, the SHE exists generically in systems with strong SOC. It has been studied \[\]in nonmagnetic ~\cite{Guo2008,Tanaka2008,Freimuth2010,Zimmermann2014} as well as antiferromagnetic~\cite{Zimmermann2014,fukami2016magnetization,oh2016field,tshitoyan2015electrical,zhang2014spin} metals.
Very recently, a strong SHE was experimentally discovered in another chiral AFM compound
Mn$_3$Ir~\cite{Zhang2016Mn3Ir}. Therefore, chiral AFM materials are appealing
candidates for finding significant AHE and SHE. They have also stimulated the search for  Weyl points in the same family of materials~\cite{Yang2016} and exotic magneto-optical Kerr effect~\cite{Feng2015}.

In this work, we have performed a comprehensive study of the intrinsic AHE and SHE of the compounds Mn$_3X$ ($X=$ Ge, Sn, Ga, Ir, Rh and Pt), using \textit{ab initio} Berry phase calculations.  These compounds
exhibit a chiral AFM order well above room temperature (see Table I).
This article is organized as follows. We first introduce the \textit{ab initio}
method and the linear-response method that we have used to compute the AHE and
SHE in Sec.~\ref{methods}. We then discuss the relationship of the symmetry of
the crystal lattice and magnetic lattice to the SHC and AHC in
Sec.~\ref{structure}. In Sec.~\ref{results}, we discuss the results of our
calculations with the assistance of symmetry analysis, where the Mn$_3X$ compounds
are classified into two groups according to their crystal and magnetic structures. Finally, we summarize our results in Sec.~\ref{summary}.

\section{methods}\label{methods}

The anomalous Hall conductivity (AHC) and spin Hall conductivity (SHC) characterize the AHE and SHE, respectively. In addition, the spin lifetime and related spin manipulation methods are important ingredients for the SHE device applications, but these aspects are beyond the scope of the current study.
The AHC and SHC have been calculated using the Berry phase that we have determined from  \textit{ab initio} band structures.
Density-functional theory (DFT) calculations were performed for the Mn$_3$X bulk crystals with the Vienna $Ab-initio$ Simulation Package (\textsc{vasp})~\cite{kresse1996} within the generalized gradient approximation (GGA)
~\cite{perdew1996}. The SOC was included in our calculations.
The material-specific Hamiltonians were established  by projecting the DFT Bloch wave functions onto maximally localized Wannier functions (MLWFs) ~\cite{Mostofi2008}.
Based on these tight-binding Hamiltonians, that include realistic material parameters,
we have calculated the intrinsic AHC and SHC by using the Kubo formula approach
within the linear response~\cite{Xiao2010, Nagaosa2010, Sinova2015, Gradhand2012}.
The AHC ($\sigma_{\alpha \beta}$) is obtained from
\begin{widetext}
\begin{equation}
\label{AHC}
\begin{aligned}
\sigma_{\alpha \beta}= -\dfrac{e^2}{\hbar}\sum_n
  \int_{BZ}\dfrac{d^3\vec{k}}{(2\pi)^3}f_n(\vec{k})\Omega_{n}(\vec{k}),   \\
\Omega_n(\vec{k})= 2i\hbar^2 \sum_{m \ne n} \dfrac{<u_{n}(\vec{k})|\hat
  v_{\alpha}|u_{m}(\vec{k})><u_{m}(\vec{k})|\hat v_{\beta}|u_{n}(\vec{k})>}{(E_{n}(\vec{k})-E_{m}(\vec{k}))^2},
\end{aligned}
\end{equation}
\end{widetext}
where $\hat{v}_{\alpha (\beta, \gamma)}=\frac{i}{\hbar}[\hat{H},\hat{r}_{\alpha (\beta, \gamma)}]$  is
the velocity operator with $\alpha,\beta,\gamma=x,y,z$; $\hat{r}_\alpha$ is the position operator. $f_{n} (\vec{k})$ is
the Fermi-Dirac distribution. $|u_{n}(\vec{k})>$ and $E_{n}(\vec{k})$
are the eigenvector and eigenvalue of the Hamiltonian $\hat{H}(\vec{k})$, respectively.
$\Omega_n(\vec{k})$ is the Berry curvature in momentum space, and
the corresponding AHC $\sigma_{\alpha \beta}$ can be evaluated by summing
the Berry curvature over the Brillouin zone (BZ) for all the occupied bands.
 Here $\sigma_{\alpha \beta}$ corresponds to a $3\times3$ matrix and
indicates a transverse Hall current $j_\alpha$ generated by a longitudinal electric field $E_\beta$, which satisfies
$J_\alpha=\sigma_{\alpha \beta}E_\beta$. For the evaluation of the velocity operator we assume for simplicity that the position operator is diagonal in the Wannier basis, as is commonly done in tight-binding calculations.

The intrinsic SHC can be obtained by replacing the velocity operator with the spin current operator
$\hat J_{\alpha}^{\gamma}=\dfrac{1}{2} \{ \hat v_{\alpha},\hat s_{\gamma}\}$,
where $\hat s_{\gamma}$ is the spin operator. The SHC then has the form of

\begin{widetext}
\begin{equation}
\label{SHC}
\begin{aligned}
\sigma_{\alpha \beta}^{\gamma}= \dfrac{e}{\hbar}\sum_n
\int_{BZ}\dfrac{d^3\vec{k}}{(2\pi)^3}f_n(\vec{k})\Omega_{n,{\alpha \beta}}^{\gamma}(\vec{k}), \\
\Omega_{n,\alpha \beta}^{\gamma}(\vec{k})= 2i \hbar^2 \sum_{m \ne n} \dfrac{<u_{n}(\vec{k})|\hat
J_{\alpha}^{\gamma}|u_{m}(\vec{k})><u_{m}(\vec{k})|\hat{v}_{\beta}|u_{n}(\vec{k})>}{(E_{n}(\vec{k})-E_{m}(\vec{k}))^2},
\end{aligned}
\end{equation}
\end{widetext}
$\Omega_{n,\alpha \beta}^{\gamma}(\vec{k})$ is referred to as the spin Berry curvature
in the following, in order to distinguish it from the Berry curvature $\vec \Omega_n(\vec{k})$.
The SHC ($\sigma_{\alpha \beta}^{\gamma}$; $\alpha,\beta,\gamma=x,y,z$)
is a third-order tensor ($3\times3\times3$) and represents the spin current ${Js}_{\alpha}^{\gamma}$
generated by an electric field $\vec{E}$ via
${Js}_{\alpha}^{\gamma}=\sigma_{\alpha \beta}^{\gamma}{E}_{\beta}$,
where ${Js}_{\alpha}^{\gamma}$ flows along the $\alpha$-direction with the
spin-polarization along the $\gamma-$direction and $E_\beta$ is the
$\beta-$component of the electric field $\vec{E}$.

For the integrals of Eqns.~\ref{AHC}-\ref{SHC}, the BZ was sampled by $k$-grids
from $50\times50\times50$ to $200\times200\times200$. Satisfactory convergence
was achieved for a $k$-grid of size $150\times150\times150$. Increasing the grid size to $200\times200\times200$ varied the SHC and AHC by no more than 5\%.
Note that the unit of SHC differs from that of the AHC by $\frac{\hbar}{2e}$, where $\hbar/2$ is the spin angular momentum and $e$ is the electron charge. Thus, the unit of SHC is $(\hbar/e)(\Omega\cdot cm)^{-1}$.

Since AHC and SHC are determined directly by the band structure, they are fully
compatible with the symmetry of the Hamiltonian. Therefore, we can use symmetry
analysis to simplify the shape of the AHC and SHC tensor matrices, by forcing
certain matrix elements to be zero and constraining some to be the same.
Here, we obtain the shape of the intrinsic response tensor from the Linear-Response-Symmetry
code~\cite{vzelezny2016spin,symcode}, which analyzes the symmetry operations of the corresponding crystal and magnetic space groups~\cite{isotropy} and then determines the tensor shape by solving the linear equations. We note that a similar study~\cite{Seemann2015} also recently considered how the shape of the tensor response varied according to the magnetic Laue group.
The shape of the AHC and SHC tensors are shown in Table II.  These are very helpful in checking the validity and numerical convergence of our calculations by comparing the symmetry of the calculated matrices and the ideal symmetry-imposed matrices.
Furthermore, the tensor shape surely relies on the coordinate system that is specified in the next section.
The AHC and SHC tensors can be expressed in different coordinate systems, which are physically
equivalent, and can be transformed into each other according to specific rotation matrices~\cite{Kleiner1966}.

\section{Crystallographic and magnetic structures}\label{structure}

The compounds considered here can be classified into two groups according to their crystallographic structure. Mn$_3$Ga, Mn$_3$Ge and Mn$_3$Sn display an hexagonal lattice with the space group $P6_3/mmc$ (No. 194).
The primitive unit cell includes two Mn$_3X$ planes that are stacked along the $c$ axis according to ``--AB--AB--''.
Each structure contains plane of Mn atoms that constitute a Kagome-type lattice with Ga, Ge or Sn lying at the center of a hexagon formed from the Mn atoms.
In the Kagome plane due to magneto-geometrical frustration, the Mn magnetic
moments exhibit a non-collinear AFM order, where the neighboring moments are
aligned at a 120$^\circ$ angle~\cite{kren1970neutron,Kadar1971, Ohoyama1961}.
The energetically favored AFM configuration was revealed, as illustrated in Fig. 2a, in earlier DFT calculations~\cite{zhang2013}.  The magnetic ordering
temperatures are above 365 K for all these three compounds, as shown in Table I.
Additionally, Mn$_3$Ga and Mn$_3$Ge can also crystallize into a tetragonal phase with a ferrimagnetic structure~\cite{kren1970neutron,Kadar1971,Balke2007}, which is not considered in this work.

\begin{figure}[htbp]
\begin{center}
\includegraphics[width=0.45\textwidth]{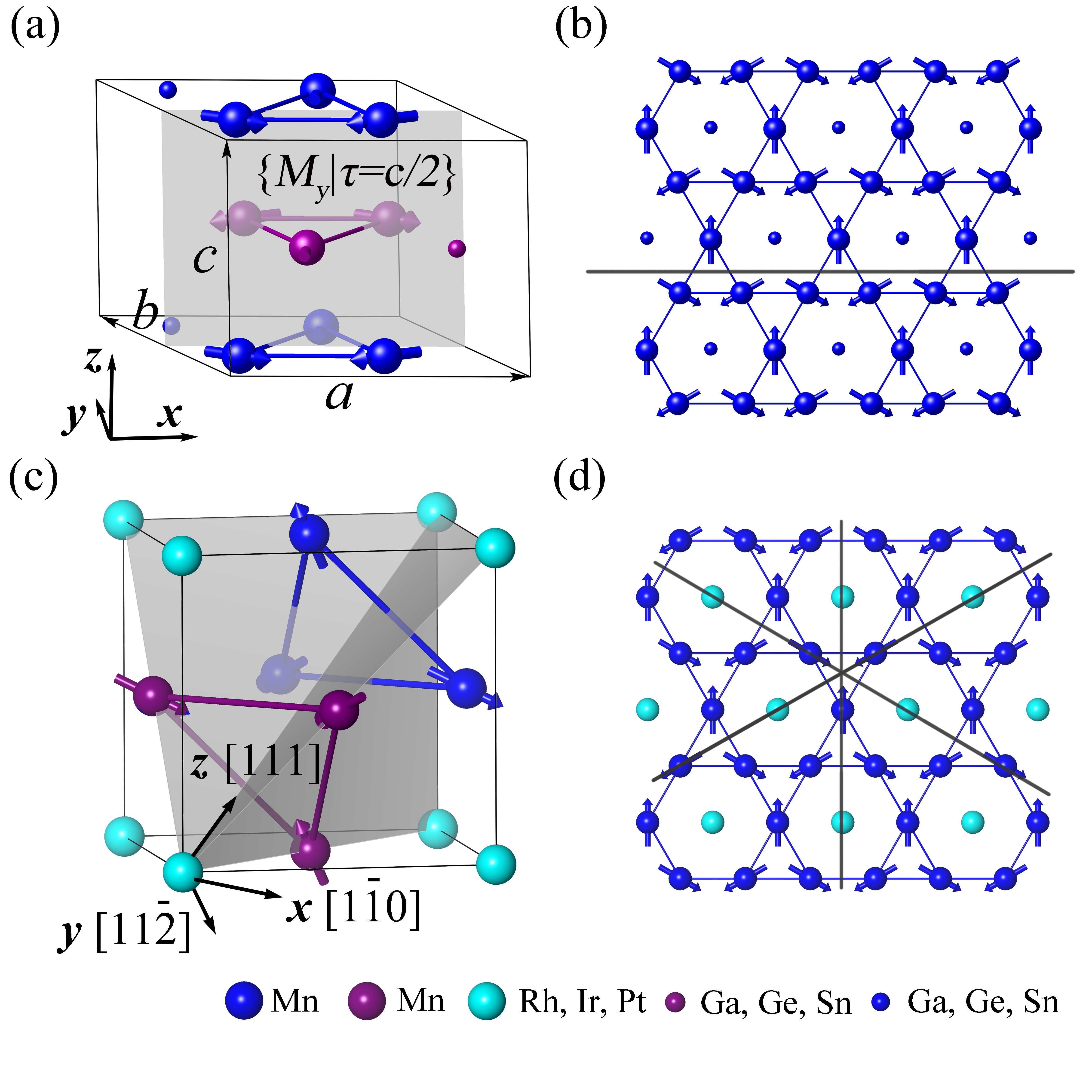}
\end{center}
\caption{Crystal lattice and magnetic structure for Mn$_3X$ ($X$ = Ga,
Ge, Sn, Rh, Ir, and Pt).
(a) Mn$_3X$ ($X$ = Ga, Ge and Sn) display
an hexagon lattice with Mn forming a Kagome sublattice stacking along the
$c$ axis. The gray plane indicates the $M_y$ mirror plane of the symmetry operation
$\{M_y | \tau = c/2 \}$.
The crystallographic $a$ and $c$ axes align with the $x$ and $z$ directions, respectively, with
the $b$ axis lying inside the $xy$ plane.
(b) Top view of the Mn Kagome-type lattice showing triangular and hexagonal arrangements of the Mn moments.
Arrows represent the Mn magnetic moments, presenting a non-collinear AFM configuration.
The mirror plane position is indicated by a black line.
(c) The crystal structure of Mn$_3X$ ($X$ = Rh, Ir and Pt) has an FCC
lattice. Three mirror planes 
are shown in gray.
Here a mirror reflection combined with a time-reversal symmetry operation preserves the magnetic lattice.
(d) The Mn sublattice also forms a Kagome-type configuration in Mn$_3X$ ($X$ = Rh, Ir and Pt), thereby forming
a non-collinear AFM phase, but which is slightly different from the Mn$_3$Ge family.
The projections of three mirror planes are indicated by black lines.
To match the hexagonal lattice conveniently, the Kagome plane is set as the $xy$ plane and the plane normal as the $z$ axis. Here $x$ is along the crystallographic [$1\bar{1}0$], $y$ along [$11\bar{2}$], and $z$ along [111].
}
\label{stru}
\end{figure}

Mn$_3$Rh, Mn$_3$Ir and Mn$_3$Pt crystallize in a
face-centered cubic (FCC) lattice (space group $Pm\bar{3}m$, No. 221) with Ir (Rh, Pt) and Mn
located at the corner and face-center sites, respectively,
as shown in Fig. 2b. Within the (111) plane, the Mn sublattice
also form a Kagome lattice. In contrast to that of Mn$_3$Ge, the Kagome planes
stack in an ``--ABC--ABC--'' sequence.
The non-collinear AFM structure
has also been observed by neutron diffraction measurements
~\cite{kren1968magnetic,tomeno1999magnetic,kren1966magnetic}.
Distinct from Mn$_3$Ge, here the magnetic moments all point towards or away from the center of the Mn triangles. The AFM order also persists to well above room temperature. (see Table I).

\begin{table}[htbp]
\centering
\caption{Crystal structure, magnetic structure and AFM ordering temperature ($T_N$) for Mn$_3X$ compounds.}
\begin{tabular}{lccc}
\hline
         &                        & Crystal           & Magnetic \\
         &   $T_N$(K)     & space group  & space group \\
\hline
Mn$_3$Ga$^a$ & 470                               & \multirow{3}{*}{$P6_3/mmc$, no. 194}     & \multirow{3}{*}{$R\overline{3}m'$}                     \\ 
Mn$_3$Ge$^b$ & 365                  &              &                                           \\ 
Mn$_3$Sn$^c$ & 420                              &  &                                           \\
\hline
Mn$_3$Rh$^d$ & $853\pm 10$                          & \multirow{3}{*}{$Pm\bar{3}m$, no. 221}   & \multirow{3}{*}{$Am'm'm2$}                    \\ 
Mn$_3$Ir$^e$ & $960\pm 10 $              &                  &                                           \\ 
Mn$_3$Pt$^f$ & $473\pm 10 $         &                       &                                           \\
\hline
$a$, Ref.~\onlinecite{kren1970neutron} &
$b$, Ref.~\onlinecite{yamada1988magnetic} &
$c$, Ref.~\onlinecite{sticht1989non} &  \\
$d$, Ref.~\onlinecite{kren1966magnetic} &
$e$, Ref.~\onlinecite{tomeno1999magnetic} &
$f$, Refs.~\onlinecite{kren1966magnetic,kren1968magnetic}\\
\end{tabular}
\label{transition}
\end{table}

\section{Results and Discussions}\label{results}

\subsection{Anomalous Hall Effect in Mn$_3$Ga, Mn$_3$Ge and Mn$_3$Sn}\label{Mn3Ge}


\begin{figure}[htbp]
\begin{center}
\includegraphics[width=0.45\textwidth]{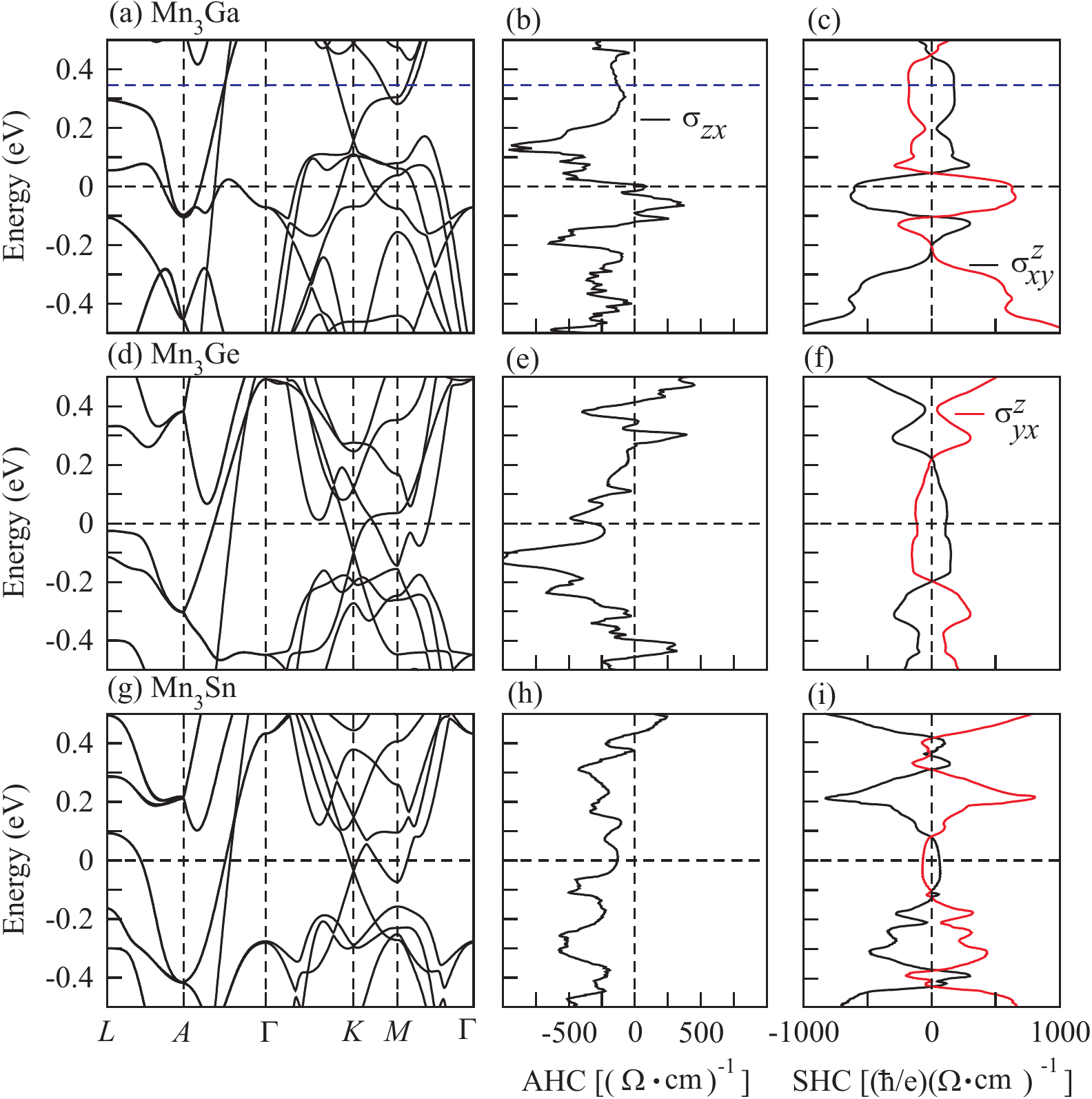}
\end{center}
\caption{
Electronic band structure for (a) Mn$_3$Ga, (d) Mn$_3$Ge, and (g) Mn$_3$Sn.
Energy dependent AHC $\sigma_{zx}$ for (b) Mn$_3$Ga, (e) Mn$_3$Ge
and (h) Mn$_3$Sn.
Energy dependent SHC tensor elements of $\sigma_{xy}^{z}$ and $\sigma_{yx}^{z}$
for (c) Mn$_3$Ga, (f) Mn$_3$Ge, and (i) Mn$_3$Sn.
Mn$_3$Ga would have the same number of valence electrons as does Mn$_3$Ge and
Mn$_3$Sn if the Fermi level is shifted up to the blue dashed line in (a,b,c)
}
\label{band-Ge}
\end{figure}

The AHC $\sigma_{\alpha \beta}$ can be understood by a consideration of the symmetry of the magnetic structure.
As indicated in Fig. 1a there is a mirror plane $\hat{M}_y$ that is parallel to the $zx$ plane.
By combining a mirror reflection about this plane and a translation operation along the $c$ direction
$\hat{\tau}=(0,0,c/2)$, the system is imaged back onto itself with the same crystallographic and magnetic structures.
Therefore, the magnetic structure in Mn$_3$Ga, Mn$_3$Ge and Mn$_3$Sn is symmetric with respect to
the  $\{\hat{M_{y}} | \hat{\tau} \}$ symmetry operator.
The mirror operation $\hat{M}_y$ changes
the signs of $\Omega_{yz}(\vec{k})$ and $\Omega_{xy}(\vec{k})$,
but preserves $\Omega_{zx}(\vec{k})$, since $\Omega_{\alpha \beta}(\vec{k})$ is a pseudovector, just like the spin.
Accordingly, $\sigma_{yz}$ and $\sigma_{xy}$ that are parallel to the mirror plane are
transformed to $-\sigma_{yz}$ and $-\sigma_{xy}$, with respect to
the $\hat{M}_y$ reflection (the translation operation does not affect the Berry curvature).
Thus, from symmetry considerations, $\sigma_{yz}$ and $\sigma_{xy}$ must be zero, and only $\sigma_{zx}$ can be nonzero.
We therefore propose that the preferred experimental setup for maximizing AHC is
to confine the electric field within the $zx$ plane, for example, by setting the
electric current along $z$ and detecting the transverse voltage along $x$.

\begin{table*}[htbp]
\caption{
Shape of the  AHC and SHC tensors obtained from symmetry analysis
and numerical calculations for Mn$_3X$ ($X$=Ga,Ge and Sn). The calculated SHC tensor elements
are set to zero when they are smaller than 12 $(\hbar/e)(\Omega\cdot cm)^{-1}$ .
The coordinates used here are $x$ along [100], $y$ along [120], and $z$ along [001],
as presented in Fig. 1(a, b). The AHC is given in units of $(\Omega\cdot cm)^{-1}$
and the SHC in units of $(\hbar/e)(\Omega\cdot cm)^{-1}$.
}
\label{table:Ge}
\centering
\begin{tabular}{@{}ccccc@{}}
\toprule
\hline
                                                                                        & \multicolumn{1}{c}{AHC}                                                                        & \multicolumn{3}{c}{SHC}                                                                                                                                                                                                                                                                                                                                                                                                                                                                                                    \\ \midrule
                                                                                        & \multicolumn{1}{c}{$\sigma$}                                                                   & \multicolumn{1}{c}{$\underline{\sigma}^x$}                                                                                                                                                     & \multicolumn{1}{c}{$\underline{\sigma}^y$}                                                                                                                                                       & \multicolumn{1}{c}{$\underline{\sigma}^z$}                                                                 \\ \hline
%
%
%

\multicolumn{1}{c}{\begin{tabular}[c]{@{}c@{}} symmetry-imposed \\ tensor shape \\ \end{tabular}} &
$\left( \begin{array}{ccc} 0 & 0 &-\sigma_{zx} \\0 & 0&0 \\ \sigma_{zx} &0&0\end{array} \right)$ &

$\left( \begin{array}{ccc}
          0       &       0          &       0       \\
          0       &       0          & \sigma_{yz}^x \\
          0       & \sigma_{zy}^x    &       0
\end{array}  \right)$  &

$\left( \begin{array}{ccc}
          0       &       0          & \sigma_{xz}^y \\
          0       &       0          &       0       \\
    \sigma_{zx}^y &       0          &       0
\end{array}  \right)$ &

$\left( \begin{array}{ccc}
          0       & \sigma_{xy}^z    &       0       \\
    \sigma_{yx}^z &       0          &       0       \\
          0       &       0          &       0
\end{array}  \right)$ \\ 

\multicolumn{1}{c}{\begin{tabular}[c]{@{}c@{}}Mn$_3$Ga\\ \end{tabular}} &
$\left( \begin{array}{ccc} 0 &    0    & -81\\ 0 & 0 & 0 \\ 81 & 0 & 0 \end{array} \right)$ &

$\left( \begin{array}{ccc} 0 & 0 & 0 \\
0 & 0  & -14 \\ 0 & 12 & 0 \end{array} \right)$ &

$\left( \begin{array}{ccc} 0 & 0 & 15 \\
0 & 0  & 0 \\ -7 & 0 & 0 \end{array} \right)$ &

$\left( \begin{array}{ccc} 0 & -597 & 0 \\
626 & 0  & 0 \\ 0 & 0 & 0 \end{array} \right)$ \\ 

\multicolumn{1}{c}{\begin{tabular}[c]{@{}c@{}}Mn$_3$Ge\\ \end{tabular}} &
$\left( \begin{array}{ccc} 0 &    0    & 330\\ 0 & 0&0 \\ -330 & 0 & 0 \end{array} \right)$ &

$\left( \begin{array}{ccc} 0 & 0 & 0 \\
0 & 0  & -21 \\ 0 & 18 & 0 \end{array} \right)$ &

$\left( \begin{array}{ccc} 0 & 0 & 21 \\
0 & 0  & 0 \\ -18 & 0 & 0 \end{array} \right)$ &

$\left( \begin{array}{ccc} 0 & 112 & 0 \\
-115 & 0  & 0 \\ 0 & 0 & 0 \end{array} \right)$ \\ 

\multicolumn{1}{c}{\begin{tabular}[c]{@{}c@{}}Mn$_3$Sn\\ \end{tabular}} &
$\left( \begin{array}{ccc} 0 &    0    & 133\\ 0 & 0 & 0 \\ -133 & 0 & 0 \end{array} \right)$ &

$\left( \begin{array}{ccc} 0 & 0 & 0 \\
0 & 0  & -36 \\ 0 & 96 & 0 \end{array} \right)$ &

$\left( \begin{array}{ccc} 0 & 0 & 36 \\
0 & 0  & 0 \\ -96 & 0 & 0 \end{array} \right)$ &

$\left( \begin{array}{ccc} 0 & 64 & 0 \\
-68 & 0  & 0 \\ 0 & 0 & 0 \end{array} \right)$ \\ \hline

\end{tabular}
\end{table*}

Our calculations are fully consistent with the above symmetry analysis,
as shown in Table II, where only $\sigma_{zx}$ ($\sigma_{xz}=-\sigma_{zx}$) is nonzero.
The AHC of Mn$_3$Ge is as large as 330
$(\Omega\cdot cm)^{-1}$. Although Mn$_3$Sn has a stronger SOC than  Mn$_3$Ge,
its AHC is less than half that of Mn$_3$Ge. Mn$_3$Ga exhibits the smallest AHC
and, moreover, the AHC has the opposite sign to those of the Ge and Sn
compounds. This is fully consistent with recent experiments on the Ge and Sn
compounds~\cite{Nakatsuji2016,Nayak2016}, where the in-plane AHC ($\sigma_{xy}$)
is negligible compared to the out-of-plane AHC ($\sigma_{zx}$ and $\sigma_{yz}$), and Mn$_3$Sn displays a smaller AHC in magnitude than Mn$_3$Ge. We note that $\sigma_{zx}$ and $\sigma_{yz}$ may be both nonzero if a different coordinate axis is chosen or the chiral moments are rotated by an external magnetic field.

Since the intrinsic AHE originates from the electronic band structure, we analyzed
the band structure in detail to understand the differences among these three compounds.
Their calculated band structures are shown in Fig.~\ref{band-Ge}.
Since the valence electrons for Ga and Ge (Sn) are $4s^24p^1$ and $4s^24p^2$ ($5s^25p^2$),
respectively, the band structure of Mn$_3$Ga looks very similar to that of
Mn$_3$Ge (Mn$_3$Sn). The Fermi level is shifted up by 0.34 eV (equivalent to adding one electron).
Correspondingly, the shapes of the energy-dependent AHC curves for
Mn$_3$Ga and Mn$_3$Ge (Mn$_3$Sn) are also very similar. The value of $\sigma_{zx}$
in Mn$_3$Ga changes sign from negative to positive after
tuning up the Fermi level.

\begin{figure}[htbp]
\begin{center}
\includegraphics[width=0.45\textwidth]{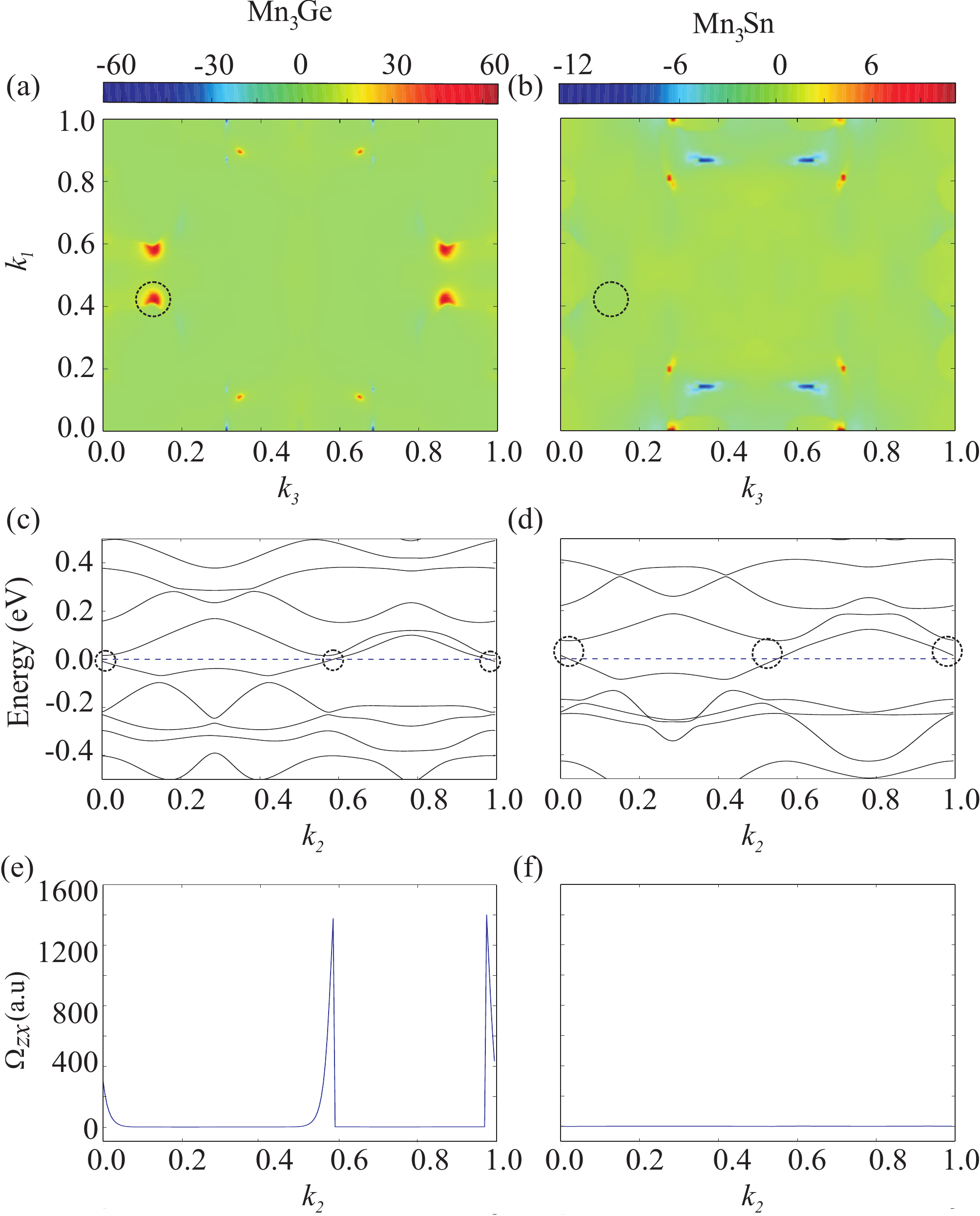}
\end{center}
\caption{
(a, b) Berry curvature projected onto the $k_3$-$k_1$ plane for Mn$_3$Ge
and Mn$_3$Sn, respectively. The Berry curvature in Mn$_3$Ge is
dominated by the four areas that are highlighted in yellow in (a). (c) Energy dispersion of
Mn$_3$Ge along $k_2$ with ($k_3$, $k_1$) fixed at the Berry curvature dominated
point (0.127, 0.428), identified as the black dashed
circle marked in (a). The band gaps are very small near $k_2$=0.5
and 1, which are noted by the small black dashed circles.
(d) Band structure of Mn$_3$Sn for the same reciprocal space cut as in (c).
The band gaps are much larger in Mn$_3$Sn, as denoted by the larger black dashed circles.
(e) The evolution of the Berry curvature ${\Omega}^y$ of Mn$_3$Ge corresponding
to the band structure given in (c). The small band gaps around $k_2$=0.5
and 1 make larger contributions to the Berry curvature.
(f) The magnitude of Berry curvature along the same path in Mn$_3$Sn is negligibly small compared to that in Mn$_3$Ge.
}
\label{ahc-zx}
\end{figure}


Atomic Ge and Sn have similar valence electron configurations while Sn has
a larger atomic radius and stronger SOC compared to Ge. Although the consequent changes in the band structures are subtle (see Fig.~\ref{band-Ge}),
the effect on the resultant AHC can be significant.
To better understand the AHE in Mn$_3$Ge and Mn$_3$Sn, we considered the
distributions of the Berry curvature in the reciprocal space. We have projected the Berry
curvature components of ${\Omega}_{zx}$ onto the $k_3$-$k_1$ ($k_z$-$k_x$) plane
by integrating them along $k_2$, where $k_{1,2,3}$ are the reciprocal lattice vectors, and $k_3$ and $k_1$ are aligned with the $k_z$ and $k_x$ axes, respectively.
The projected Berry curvatures
of Mn$_3$Ge and Mn$_3$Sn with the Fermi level lying at the charge
neutral point are shown in Figs.~\ref{ahc-zx}(a) and ~\ref{ahc-zx}(b), respectively.
One can easily identify the origin of the significant differences of the Berry curvature between Mn$_3$Ge
and Mn$_3$Sn.
The large AHC mainly arises from
the positive $hot~spots$ located around (0.127,
0.428) (the coordinates are in units of the reciprocal lattice vectors $k_1$ and $k_2$)
and its three partners in the
$k_3$-$k_1$ plane, while these four hot spots are not seen in Mn$_3$Sn.
Taking the hotspot at (0.127, 0.428)
as an example, we have checked the band structure and corresponding
Berry curvature evolution with $k_2$ varying from 0 to 1.
From the band structure of Mn$_3$Ge in Fig.~\ref{ahc-zx}(c) we can see
that the Fermi level crosses two small gaps around $k_2$=0 and 0.5. According to Eqn. 1, the entanglement
between occupied and unoccupied states must be very strong
around these two points and contributes to a large Berry curvatures, as indicated by the two peaks in Fig.~\ref{ahc-zx}(e).
This is fully consistent with previous calculations on Mn$_3$Ge~\cite{Nayak2016}.
Mn$_3$Sn has a similar band structure along the
same $k$-path, as can be seen by comparing Fig.~\ref{ahc-zx}(c) and (d), whereas
the band gaps around $k_2$=0 and 0.5 are much larger compared
to that in Mn$_3$Ge. Consequently, the two Berry curvature peaks
 disappear in Mn$_3$Sn, as shown in Fig.~\ref{ahc-zx}(f). Thus, a tiny
changes in band structure can result in significant changes in the
Berry curvature and AHC in this class of compounds.

\begin{figure}[htbp]
\begin{center}
\includegraphics[width=0.4\textwidth]{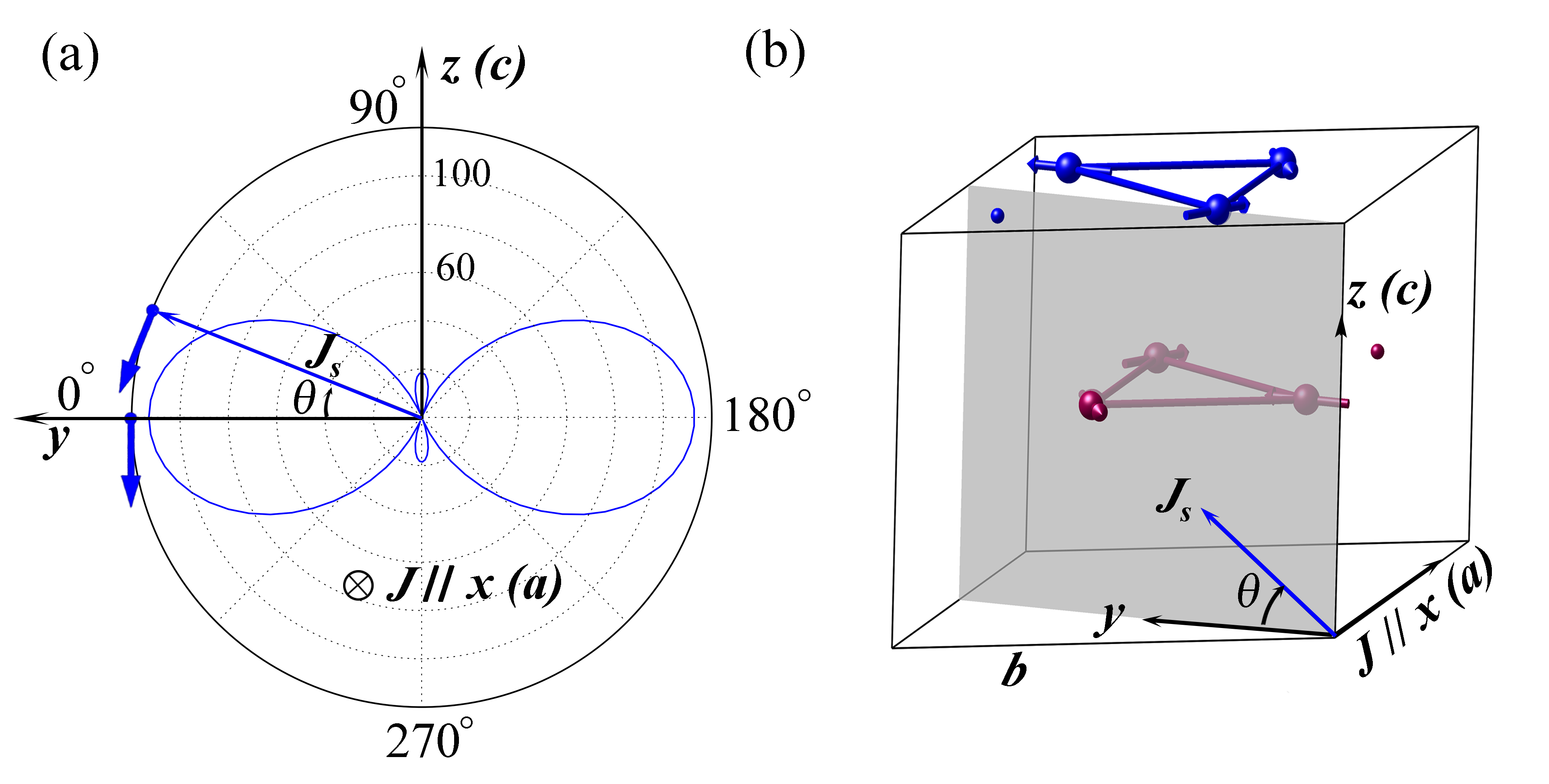}
\end{center}
\caption{
(a) Angle-dependent spin current $Js$ arising from a charge current $J$ along the $x$ axis (the crystallographic $a$ lattice vector) in Mn$_3$Ge. $Js$ rotates inside the $yz$ plane. The largest spin Hall conductivity is when $Js$ is along the $y$ axis ($\theta=0^\circ$). The blue arrows represent the spin polarization directions of $Js$.
(b) Schematic of $Js$ and $J$ with respect to the lattice orientation.
}
\label{angel-Ge}
\end{figure}

\begin{figure}[htbp]
\begin{center}
\includegraphics[width=0.45\textwidth]{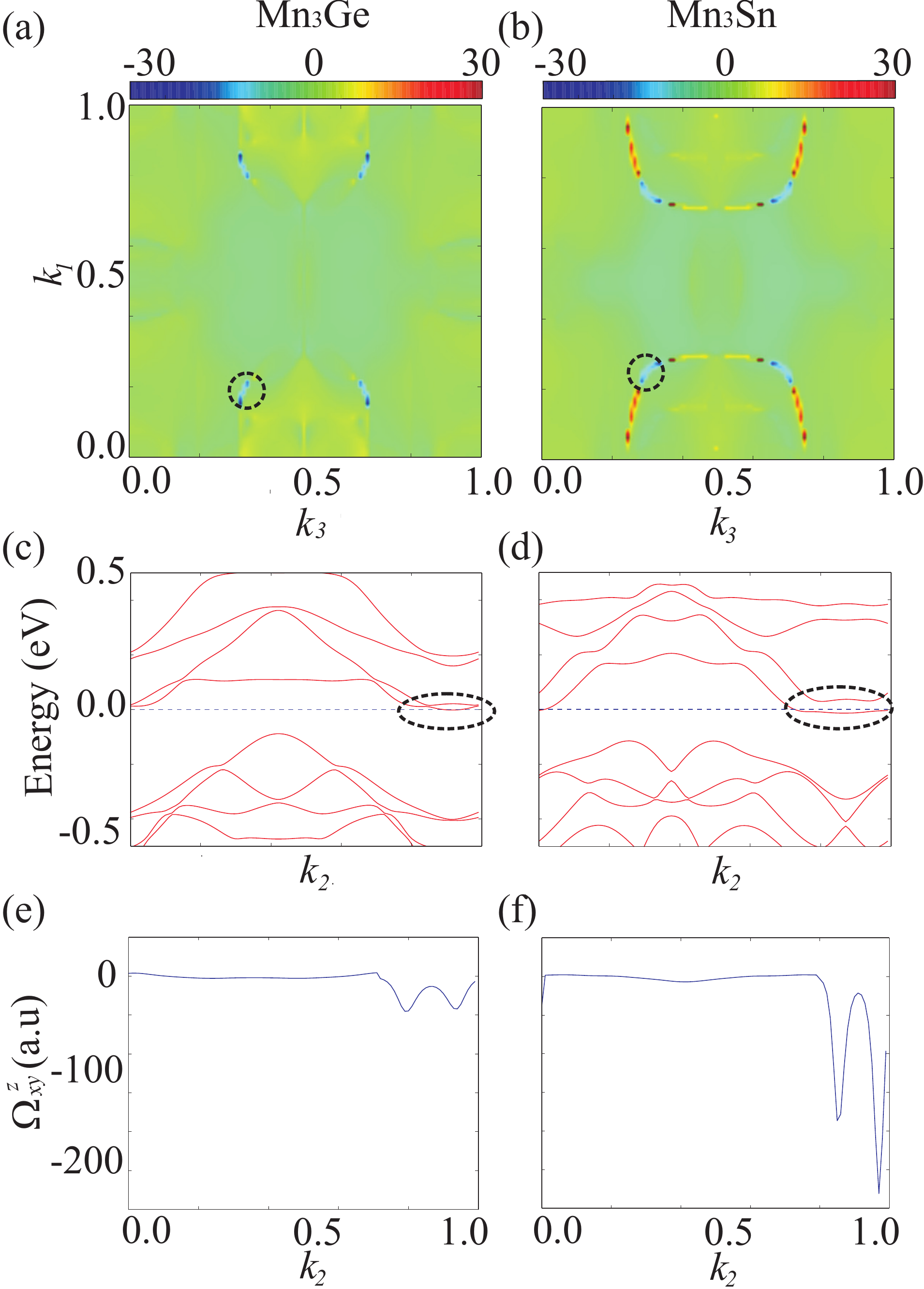}
\end{center}
\caption{
(a, b) Spin Berry curvature projected onto the $k_3$-$k_1$ plane for Mn$_3$Ge
and Mn$_3$Sn, respectively. The two compounds have similar distributions
of the projected spin Berry curvature. (c, d) Band structures of Mn$_3$Ge
and Mn$_3$Sn, respectively, along $k_2$. The coordinates of $k_3$ and $k_1$
are fixed at (0.31, 0.15) and
(0.29, 0.24) for Mn$_3$Ge
and Mn$_3$Sn, respectively, as noted by the black dashed circles in
(a) and (b). Small band gaps exist around the $k_2$=1
point for both Mn$_3$Ge and Mn$_3$Sn, as marked by the black dashed circles.
(e, f) Corresponding spin Berry curvature evolutions along $k_2$ for
Mn$_3$Ge and Mn$_3$Sn, respectively.  The large spin Berry curvature mainly
originates from the small band gaps in the band structures.
}
\label{shc-zx}
\end{figure}

\subsection{Spin Hall Effect in Mn$_3$Ga, Mn$_3$Ge and Mn$_3$Sn}\label{Mn3Ge}

By adding the spin degree of freedom, the SHC becomes a third-order
tensor. Similar to the AHC, some SHC tensor elements will be exactly zero or will
be identical based on the corresponding lattice and magnetic symmetries.
The magnetic space group for Mn$_3X$ ($X$=Ga, Ge and Sn) is identified to be
$R\overline{3}m'$ and the corresponding Laue group is
$m'm'm'$~\cite{isotropy}.
The calculated shape of the intrinsic SHC tensor and corresponding numerical results
are presented in Table II. Furthermore, the SHC of Mn$_3X$ ($X$=Ga, Ge and Sn)
is strongly anisotropic with dominant components of $\sigma_{xy}^z$
and $\sigma_{yx}^z$.  These results will provides helpful information for the
experimental detection of the SHE. To illustrate the anisotropy of the SHC, we show the angle-dependent
SHC for Mn$_3$Ge in Fig.~\ref{angel-Ge}. When the charge current $J$ is fixed along $x$-axis ($a$ direction)
 and by considering the spin current $Js$ perpendicular to $J$ and rotating it,
 the corresponding magnitude of SHC is maximal for $Js || y$ while being zero for  $Js || z$.
Therefore, to observe large SHC, one should set the charge current and spin current inside the Kagome ($xy$)
plane, for example with the electron current set along $x$ and by measuring the transverse spin current
along $y$ with its spin polarization along $z$. Therefore, we stress that for optimizing the efficiency of devices that rely on
SHE and AHE, the direction of the charge current and the resulting spin current will depend on the respective compound.

As shown in Table II, the largest SHCs $\sigma_{xy}^z$ and $\sigma_{yx}^z$ are
of the order of 120 $(\hbar/e)(\Omega\cdot cm)^{-1}$ in magnitude for Mn$_3$Ge.
With the relatively small electrical conductivity (about 3300 $(\Omega\cdot
cm)^{-1}$), we would have a spin Hall angle up to $3\%$. Also the $\sigma_{yx}^z$
elements in Mn$_3$Ga is around 600 $(\hbar/e)(\Omega\cdot cm)^{-1}$.
Additionally, it is not surprising that $\sigma_{xy}^z$ and $\sigma_{yx}^z$ are not equal in magnitude,
for the $x$ and $y$ directions are not equivalent in a Kagome structure.

Since the SHC is strongly related to the location of the Fermi level, the SHC
varies quickly as the Fermi energy is shifted, especially for the metallic band structures shown in Fig.~\ref{band-Ge}.
The energy-dependent SHC of the most prominent tensor elements $\sigma_{xy}^z$ and
$\sigma_{yx}^z$  for the three compounds are shown in  Figs.~\ref{band-Ge} (e),
(f) and (i). Owing to the similar crystal lattice
constant and the same magnetic order, the shapes of the SHC curves are very similar, if we ignore the fact that Ga has one electron less than either Ge or Sn.
For Mn$_3$Ga, the SHC exhibits a minimum at the Fermi level, the charge neutral point, and increases quickly if the Fermi level moves up or down. Hence an even larger SHC is expected for Mn$_3$Ga with small electron or hole doping.
One can see that the SHC remains relatively stable with respect to varying the Fermi level in the energy window of $\pm$ 0.1 eV
 for Mn$_3$Ge and Mn$_3$Sn. Thus indicates that the SHC in the Ge and Sn compounds is robust.

Since the spin Berry curvature is distinct from the Berry curvature, the SHC and AHC can have dominant contributions from different electronic bands. Although Mn$_3$Ge and Mn$_3$Sn display very different AHCs in magnitude, their SHCs are very close.
Therefore, we expect a similar spin Berry curvature distribution in $k$ space for both compounds.
Taking the components of $\Omega_{xy}^z$ as an example, we compare the
spin Berry curvature distributions for Mn$_3$Ge and Mn$_3$Sn with the Fermi
energy lying at the charge neutral point. Similar to the above analysis
for the AHE, we also project the spin Berry curvature onto the
$k_3-k_1$ plane by integrating $\Omega_{xy}^z$ along $k_2$. As shown
in Fig.~\ref{shc-zx}, Mn$_3$Ge and Mn$_3$Sn display similar features in their respective
spin Berry curvature distributions. The shapes of the dominant areas
are very similar in both compounds, with just a little shift
within the $k_3-k_1$ plane. The dominant contribution forms thick
arcs with a transition point between positive and negative amplitudes, where the integrated spin
Berry curvature transfers from positive to negative. Since the size
of the positive dominant area is much larger than that of the negative part, the
integral of the spin Berry curvature in the whole BZ gives a positive
SHC $\sigma_{xy}^z$, as is listed in Table II.


The above positive-negative spin Berry curvature distribution is reminiscent of the similar feature of the
SHE around Weyl point, where positive and negative spin Berry curvature appear with the Weyl point as the
transition point~\cite{Yang2016}. In fact Weyl points also exist in Mn$_3$Ge and Mn$_3$Sn, however, the spin Berry
curvature transition point in Fig.~\ref{shc-zx} does not exactly overlap with the Weyl point.  A careful inspection
of the band dispersions along $k_2$ through these hot spots reveals tiny band gaps that contribute to the peaks of
the spin Berry curvature, as shown in Figs. ~\ref{shc-zx} (e) and (f).
Therefore, the intrinsic SHC mainly arises from the small band gaps lying very close to the Fermi level.

\subsection{Anomalous Hall effect and spin Hall effect in Mn$_3$Rh, Mn$_3$Ir and Mn$_3$Pt}\label{Mn3Ir}

\begin{figure}[htbp]
\begin{center}
\includegraphics[width=0.45\textwidth]{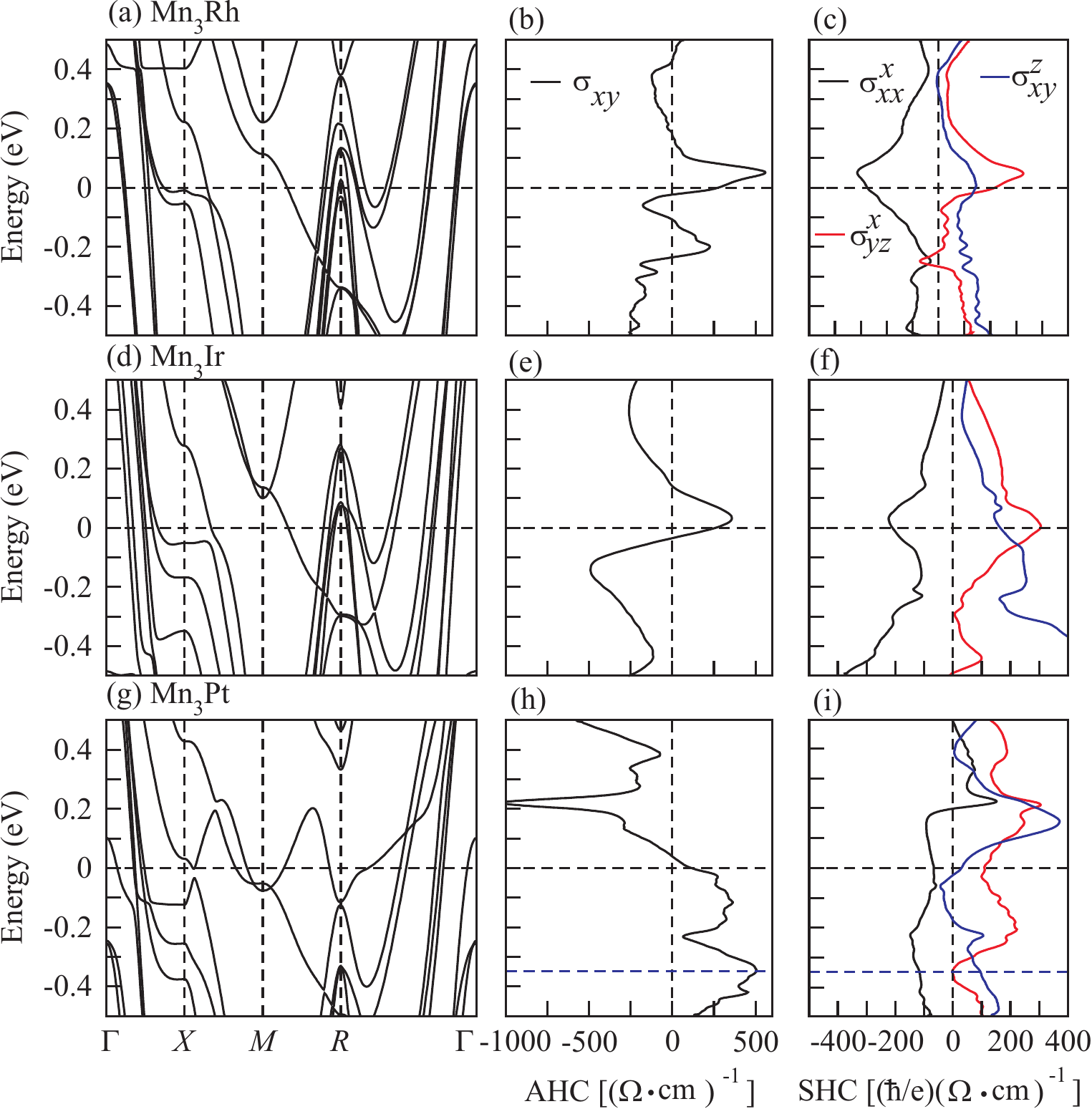}
\end{center}
\caption{
Electronic band structure for (a) Mn$_3$Rh, (d) Mn$_3$Ir, and (g) Mn$_3$Pt.
Energy dependent AHC element of $\sigma^{z}$ for (b) Mn$_3$Rh, (e) Mn$_3$Ir
and (h) Mn$_3$Pt. Energy dependent SHC tensor elements
$\sigma_{xx}^{x}$, $\sigma_{yz}^{x}$ and $\sigma_{xy}^{z}$ for
(c) Mn$_3$Rh, (f) Mn$_3$Ir, and (i) Mn$_3$Pt.
Mn$_3$Pt would have the same number of valence electrons as Mn$_3$Rh and
Mn$_3$Ir if the Fermi level shifts down to the blue dashed line in (g,h,i)
}
\label{band-Ir}
\end{figure}

\begin{figure}[htbp]
\begin{center}
\includegraphics[width=0.40\textwidth]{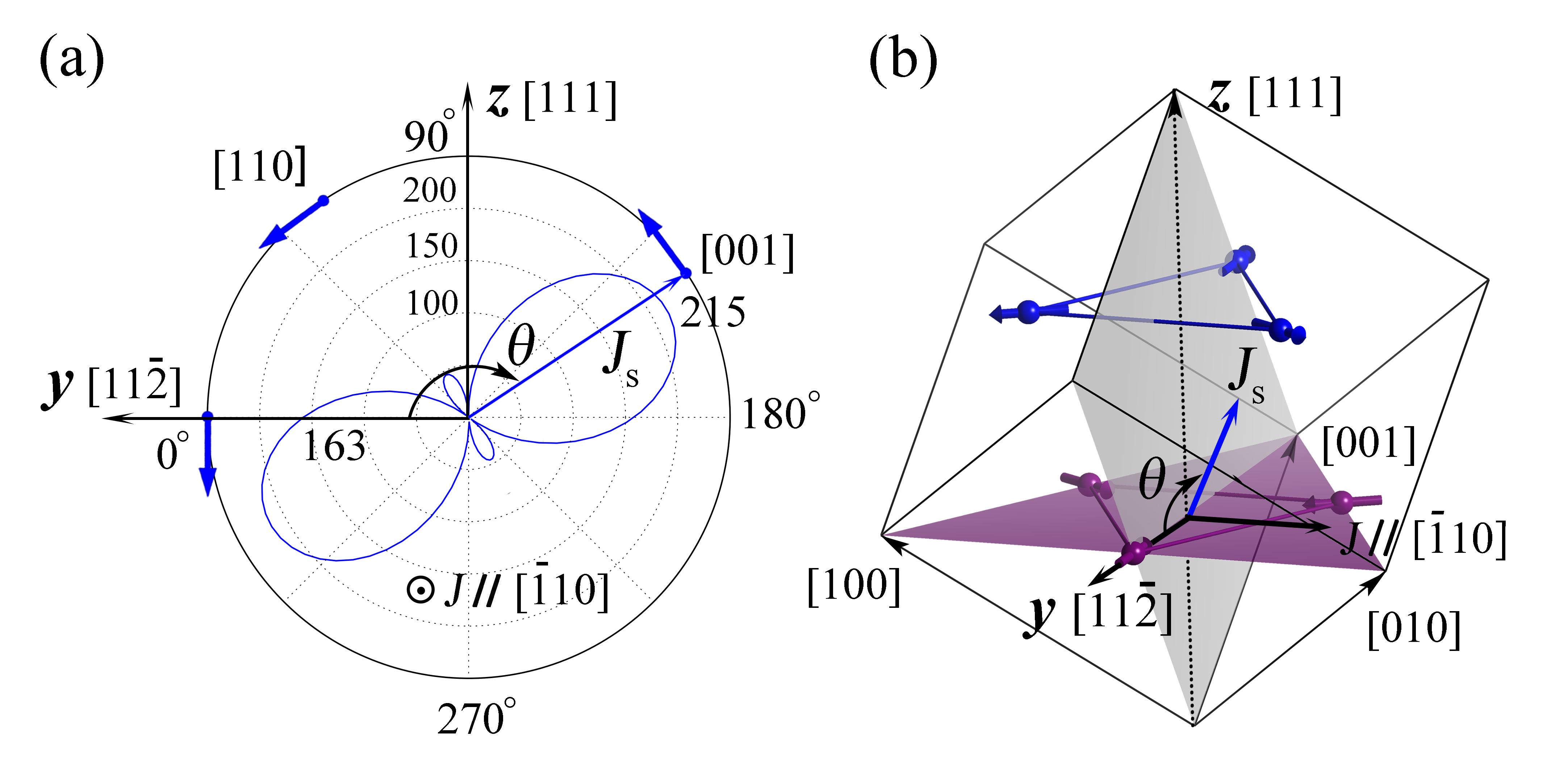}
\end{center}
\caption{
(a) Angle-dependent spin Hall conductivity for Mn$_3$Ir.
The charge current $J$ flows along $x$ (i.e. [$1\bar{1}0$]) and the resulting spin current $Js$ is shown in the $yz$ plane.
The SHC is largest when $Js || [001]$ and smallest when $Js || [111]$.
(b) Schematic of $J$  and $Js$ within the crystal structure.
}
\label{angle-Ir}
\end{figure}

\begin{table*}[htbp]
\label{table:Ir}
\caption{
The intrinsic AHC and SHC tensors obtained from symmetry analysis
and numerical calculations for Mn$_3X$ ($X$= Rh, Ir and Pt).
The calculated SHC tensor elements
are set to zero when they are smaller than 5 $(\hbar/e)(\Omega\cdot cm)^{-1}$.
The coordinate axes correspond to $z$ oriented in the [111] direction, $x$ along $[1\overline{1}0]$
and $y$ along $[11\overline{2}]$. The AHC is in units of $(\Omega\cdot cm)^{-1}$
and the SHC is in units of $(\hbar/e)(\Omega\cdot cm)^{-1}$.
}
\centering
\begin{tabular}{@{}ccccc@{}}
\toprule \hline
                                                                                        & \multicolumn{1}{c}{AHC}                                                                        & \multicolumn{3}{c}{SHC}                                                                                                                                                                                                                                                                                                                                                                                                                                                                                                    \\ \midrule
                                                                                        & \multicolumn{1}{c}{$\sigma$}                                                                   & \multicolumn{1}{c}{$\underline{\sigma}^x$}                                                                                                                                                     & \multicolumn{1}{c}{$\underline{\sigma}^y$}                                                                                                                                                       & \multicolumn{1}{c}{$\underline{\sigma}^z$}                                                                       \\ \hline
%
%
%

\multicolumn{1}{c}{\begin{tabular}[c]{@{}c@{}} symmetry-imposed \\ tensor shape  \end{tabular}} &
$\left( \begin{array}{ccc} 0 & \sigma_{xy} & 0 \\-\sigma_{xy}& 0&0 \\0&0&0\end{array} \right)$ &

$\left( \begin{array}{ccc} \sigma_{xx}^x & 0 & 0 \\
 0 & -\sigma_{xx}^x & \sigma_{yz}^x \\ 0 & \sigma_{zy}^x & 0 \end{array} \right)$ &

$\left( \begin{array}{ccc} 0 & -\sigma_{xx}^x & -\sigma_{yz}^x \\ -\sigma_{xx}^x &
 0  & 0 \\ -\sigma_{zy}^x & 0 & 0 \end{array} \right)$ &

$\left( \begin{array}{ccc} 0 & \sigma_{xy}^z & 0 \\-\sigma_{xy}^z & 0 &
0 \\ 0 & 0 & 0 \end{array} \right)$ \\  

\multicolumn{1}{c}{\begin{tabular}[c]{@{}c@{}}Mn$_3$Rh\\\end{tabular}} &
$\left( \begin{array}{ccc} 0 &    -284  & 0 \\ 284 & 0&0 \\ 0 & 0 & 0 \end{array} \right)$ &

$\left( \begin{array}{ccc} -276 & 0 & 0 \\
0 & 276  & 220 \\ 0 & 70 & 0 \end{array} \right)$ &

$\left( \begin{array}{ccc} 0 & 276 & -220 \\
276 & 0  & 0 \\ -70 & 0 & 0 \end{array} \right)$ &

$\left( \begin{array}{ccc} 0 & 145 & 0 \\
-145 & 0  & 0 \\ 0 & 0 & 0 \end{array} \right)$ \\ 

\multicolumn{1}{c}{\begin{tabular}[c]{@{}c@{}}Mn$_3$Ir\\ \end{tabular}} &
$\left( \begin{array}{ccc} 0 &    -312  & 0 \\ 312 & 0&0 \\ 0 & 0 & 0 \end{array} \right)$ &

$\left( \begin{array}{ccc} -210 & 0 & 0 \\
0 & 210  & 299 \\ 0 & -7 & 0 \end{array} \right)$ &

$\left( \begin{array}{ccc} 0 & 210 & -299 \\
210 & 0  & 0 \\ 7 & 0 & 0 \end{array} \right)$ &

$\left( \begin{array}{ccc} 0 & 163 & 0 \\
-163 & 0  & 0 \\ 0 & 0 & 0 \end{array} \right)$ \\

\multicolumn{1}{c}{\begin{tabular}[c]{@{}c@{}}Mn$_3$Pt\\ \end{tabular}} &
$\left( \begin{array}{ccc} 0 &    98  & 0 \\ -98 & 0&0 \\ 0 & 0 & 0 \end{array} \right)$ &

$\left( \begin{array}{ccc} -66 & 0 & 0 \\
0 & 66  & 108 \\ 0 & 7 & 0 \end{array} \right)$ &

$\left( \begin{array}{ccc} 0 & 66 & -108 \\
66 & 0  & 0 \\ -7 & 0 & 0 \end{array} \right)$ &

$\left( \begin{array}{ccc} 0 & 32 & 0 \\
-32 & 0  & 0 \\ 0 & 0 & 0 \end{array} \right)$ \\ \hline

\end{tabular}
\end{table*}

In the cubic lattice of Mn$_3$Rh, Mn$_3$Ir and Mn$_3$Pt,
there are three mirror planes that intersect the crystallographic [111] axis and
which are related to each other by a three-fold rotation.
The mirror reflection $\hat{M}$ preserves the lattice symmetry but reverses all spins in the Kagome plane.
Since time-reversal symmetry $\hat{T}$ can also reverse spins,
the combined symmetry of time-reversal and mirror symmetry, $\hat{T}\hat{M}$ is the symmetry of the system.
$\hat{T}\hat{M}$ forces the out-of-mirror-plane AHC components to be zero, since
the out-of-plane Berry curvature is odd with respect to $\hat{T}$ but even with respect to $\hat{M}$.
Given the existence of the three mirror planes, the only nonzero AHC component is along the co-axis of these three planes, i.e. the [111] axis.
For the convenience of the symmetry analysis, we used coordinates with $z$ along the [111] direction and
$x$, $y$ within the Kagome plane (see Fig. 1).

Our numerical calculations are again consistent with the
symmetry analysis. The AHC for Mn$_3$Ir can reach $\sigma_{xy}=312
(\Omega\cdot cm)^{-1}$ with the electric field lying in the (111) plane,
as presented in Table III, which agrees with previous calculations~\cite{Chen2014Mn3Ir}.
Compared to Mn$_3$Ir, Mn$_3$Rh exhibits similar AHC in magnitude while Mn$_3$Pt shows a much smaller AHC.
Mn$_3$Rh and Mn$_3$Ir show very similar trends in the Fermi-energy-dependent AHC, as shown in Figs.~\ref{band-Ir}(b) and (e). The peak values appear around 50 meV above the charge neutral point for both Mn$_3$Rh and Mn$_3$Ir.
Therefore, in order to get strong AHE, one simply needs weak electron
doping, and the AHC in the (111) plane can then reach 450 and 500
$(\Omega\cdot cm)^{-1}$ for Mn$_3$Rh and Mn$_3$Ir, respectively.
Compared to Rh and Ir, Pt has one more valence electron.
Hence the Mn$_3$Pt can be viewed as a strongly doped version of Mn$_3$Ir,
which shifts the Fermi level a little further beyond the peak values,
leading to a small AHC of 98 $(\Omega\cdot cm)^{-1}$ , as shown in Fig.~\ref{band-Ir}(h).

The magnetic space group for Rh, Ir and Pt compounds is $Am'm'm2$, from
which we can obtain the symmetry of the SHC tensor. As shown
in Table III, there are only four independent nonzero
elements. Our numerical calculations fit
the symmetry-imposed tensor shape very well, as shown in Table III.
The largest SHC tensor elements are $\sigma_{yz}^x$ ($\sigma_{xz}^y =\sigma_{yz}^x$) and
$\sigma_{xy}^y$ ($\sigma_{xy}^y$=$\sigma_{yx}^y$=$-\sigma_{xx}^x$=$\sigma_{yy}^x$) for Mn$_3$Rh and Mn$_3$Ir.
Therefore, the optimal experimental arrangement for large SHC is to align $Js$ within the $xy$ plane [the (111) Kagome plane]. It is interesting that $J$, $Js$ and the spin polarization of $Js$ are not necessarily perpendicular to each other and even can be parallel, as indicated by the nonzero $\sigma_{xx}^x$. The large value of $\sigma_{xx}^x$ shows a longitudinal spin current $Js$ induced by a charge current $J$ along the same direction. Such a longitudinal spin current is common in FM metals where conduction electrons are naturally spin-polarized. However, it is interesting that  these three AFM compounds can generate a similar spin current, which may promise novel spintronic applications.
In previous experiments on Mn$_3$Ir~\cite{Zhang2016Mn3Ir}, the spin current was
measured in two cases where the charge current was fixed along the [$\bar{1}10$] crystallographic direction (i.e. $x$ axis of the current work), $Js$ along the [111] (i.e. $z$ axis) direction and the [001] direction. The former case was found to exhibit a much smaller SHE than the latter one.
Therefore, we calculate the angle-dependent SHC by fixing $J||x$ and rotating $Js$ in the $yz$ plane for Mn$_3$Ir, as shown in Fig.~\ref{angle-Ir}.
One  clearly  sees that the SHC is only 7 $(\hbar/e)(\Omega\cdot cm)^{-1}$ for the former case ($\theta=0^\circ$) and
215 $(\hbar/e)(\Omega\cdot cm)^{-1}$ for the latter case ($\theta=144.7^\circ$).

Similar to the AHC, the peak values of $\sigma_{xx}^x$ and $\sigma_{yz}^x$ also appear
around 50 meV above the charge neutral point for Mn$_3$Rh and Mn$_3$Ir, while their $\sigma_{xy}^z$ is quite small.
In contrast, the Fermi-energy-dependent AHC of Mn$_3$Pt shows a similar shape to that of the Rh and Ir compounds but the corresponding Fermi energy should be upshifted by one additional electron, as shown in Fig.~\ref{band-Ir}(i). Thus, it is not surprising that $\sigma_{xy}^z$ shows a large magnitude at the charge neutral point for Mn$_3$Pt.

\section{Summary}\label{summary}

\begin{table}[htbp]
\centering
\caption{
Summary of the optimal experimental arrangements to realize large AHE and SHE.
The $xy$ plane refers to the Kagome AFM plane and the $z$ direction is perpendicular to this plane.
For AHE, the preferred plane to set the current and detect the Hall voltage is specified. For SHE,
the charge current $J$ and spin current $Js$ directions, which are supposed to be orthogonal, are indicated.
}
\begin{tabular}{lll}
\hline
                    &   AHE     & SHE  \\
\hline
Mn$_3$Ga  &  \multirow{3}{*}{$xz$ plane} &    \multirow{3}{*}{ $xy$ plane}   \\ 
Mn$_3$Ge  &                                             &                                                         \\ 
Mn$_3$Sn  &                                             &                                                         \\
\hline
Mn$_3$Rh &  \multirow{3}{*}{$xy$ plane} &  \multirow{3}{*}{$Js$ inside the $xy$ plane}     \\ 
Mn$_3$Ir   &                                             &                                                                 \\ 
Mn$_3$Pt  &                         &                              \\
\hline
\end{tabular}
\end{table}

In summary, we have studied the intrinsic AHE and
SHE in the non-collinear AFM compounds Mn$_3X$ ($X=$Ge, Sn, Ga, Rh, Ir, and Pt) by
$ab~initio$ calculations.
Large AHC and large SHC are found for these materials, which are also highly anisotropic and in agreement with recent experimental measurements.
Such an anisotropy is closely related to the symmetry of the AFM Kagome lattice, which can be helpful in rationalizing the numerical results.
Based on our calculations, we have proposed the optimal experimental setups to maximize the AHE and SHE for different systems, as shown in Table IV.
Although the SOC magnitude increases from Rh, to Ir and to Pt, the magnitude of
the corresponding AHC and SHC do not follow the same trend. This is also true
for the Ga, Ge, and Sn compounds. This indicates that the electron filling and
the detailed band structures are essential in determining the magnitude of the AHE and SHE.
We point out that the largest SHC attains a value of around 600
$(\hbar/e)(\Omega\cdot cm)^{-1}$ in Mn$_3$Ga.
Our work provides insights in the interpretation and realization of a strong AHE
and SHE in chiral AFM materials.

\begin{acknowledgments}
We thank Jario Sinova, Jereon van den Brink, and Carsten Timm for helpful discussions.
C.F. acknowledges the funding support by ERC (Advanced Grant No. 291472 "Idea
Heusler").
Y.Z. and B.Y. acknowledge the German Research Foundation (DFG) SFB 1143.
\end{acknowledgments}

%

\end{document}